\documentclass[conference]{IEEEtran}

\usepackage[utf8]{inputenc}
\usepackage[pdftex]{xcolor}
\usepackage[ruled]{algorithm}
\usepackage[noend]{algpseudocode}
\usepackage{amsmath}
\usepackage{amsfonts}
\usepackage{amssymb}
\usepackage{mathtools}
\usepackage{todonotes}
\usepackage[ligature,reserved]{semantic}
\usepackage{xspace}

\usepackage{tikz}
\usepackage{adjustbox}

\algblockdefx[Upon]{Upon}{EndUpon}%
[1]{{\bf upon} (#1) {\bf do}}%
{{\bf end upon}}

\mathlig{<-}{\leftarrow}
\mathlig{==}{\equiv}
\mathlig{<<}{\langle}
\mathlig{>>}{\rangle}
\reservestyle{\variables}{\text}
\variables{epoch,current,history,theset[the\_set],pending}

\reservestyle{\setops}{\text}
\setops{add,get,Init,Add,Get,BAdd,EpochInc,Broadcast,Deliver,GetEpoch,Propose,SetDeliver}

\reservestyle{\structs}{\text}
\structs{DPO,BAB,BRB,SBC,DSO}

\reservestyle{\stmt}{\textbf}
\stmt{call,ack,drop,return,assert,wait}

\reservestyle{\messages}{\texttt}
\messages{madd[add],mepochinc[epinc]}

\reservestyle{\api}{\texttt}
\api{apiBAdd,apiAdd,apiGet,apiEpochInc,apiTheSet[{the\_set}],apiHistory}

\reservestyle{\schain}{\texttt}
\schain{history,epoch,theset[{the\_set}],add,get,epochinc[{epoch\_inc}],getepoch[{get\_epoch}]}

\newtheorem{property}{Property}

\newtheorem{lemma}{Lemma}

\usepackage{xspace}
\usepackage{url}
\usepackage{paralist}

\newcommand{\PR}[1]{\ensuremath{\textit{#1}}}
\newcommand{\GS}{\PR{GS}\xspace}
\newcommand{\EPOCH}{\<epoch>}

\newcommand{\HISTORY}{\<history>}
\newcommand{\THESET}{\<theset>}
\newcommand{\epochs}{Setchain\xspace}

\newcommand{\BRBdeliver}{\<BRB>.\<Deliver>}
\newcommand{\BRBbroadcast}{\<BRB>.\<Broadcast>}
\newcommand{\BABdeliver}{\<BAB>.\<Deliver>}
\newcommand{\BABbroadcast}{\<BAB>.\<Broadcast>}
\newcommand{\setchain}{Setchain\xspace}
\newcommand{\setchains}{Setchains\xspace}

\newcommand{\Hspeed}{H1}
\newcommand{\Hbetteralg}{H2}
\newcommand{\Hagr}{H3}
\newcommand{\Hbyzantine}{H4}
\newcommand{\Hdegrade}{H5}

\newcommand{\Propset}{\PR{propset}}
\newcommand{\Servers}{N}

\usepackage{listings, xcolor}

\definecolor{verylightgray}{rgb}{.97,.97,.97}

\lstdefinelanguage{Solidity}{
	keywords=[1]{anonymous, assembly, assert, balance, break, call, callcode, case, catch, class, constant, continue, constructor, contract, debugger, default, delegatecall, delete, do, else, emit, event, experimental, export, external, false, finally, for, function, gas, if, implements, import, in, indexed, instanceof, interface, internal, is, length, library, log0, log1, log2, log3, log4, memory, modifier, new, payable, pragma, private, protected, public, pure, push, require, return, returns, revert, selfdestruct, send, solidity, storage, struct, suicide, super, switch, then, this, throw, transfer, true, try, typeof, using, value, view, while, with, addmod, ecrecover, keccak256, mulmod, ripemd160, sha256, sha3}, 
	keywordstyle=[1]\color{blue}\bfseries,
	keywords=[2]{set,address, bool, byte, bytes, bytes1, bytes2, bytes3, bytes4, bytes5, bytes6, bytes7, bytes8, bytes9, bytes10, bytes11, bytes12, bytes13, bytes14, bytes15, bytes16, bytes17, bytes18, bytes19, bytes20, bytes21, bytes22, bytes23, bytes24, bytes25, bytes26, bytes27, bytes28, bytes29, bytes30, bytes31, bytes32, enum, int, int8, int16, int24, int32, int40, int48, int56, int64, int72, int80, int88, int96, int104, int112, int120, int128, int136, int144, int152, int160, int168, int176, int184, int192, int200, int208, int216, int224, int232, int240, int248, int256, mapping, string, elem, uint, uint8, uint16, uint24, uint32, uint40, uint48, uint56, uint64, uint72, uint80, uint88, uint96, uint104, uint112, uint120, uint128, uint136, uint144, uint152, uint160, uint168, uint176, uint184, uint192, uint200, uint208, uint216, uint224, uint232, uint240, uint248, uint256, var, void, ether, finney, szabo, wei, days, hours, minutes, seconds, weeks, years},	
	keywordstyle=[2]\color{teal}\bfseries,
	keywords=[3]{block, blockhash, coinbase, difficulty, gaslimit, number, timestamp, msg, gas, sender, sig, value, now, tx, gasprice, origin, add, epochinc, get, setminus, emptyset},	
	keywordstyle=[3]\color{violet}\bfseries,
	identifierstyle=\color{black},
	sensitive=false,
	comment=[l]{//},
	morecomment=[s]{/*}{*/},
	commentstyle=\color{gray}\ttfamily,
	stringstyle=\color{red}\ttfamily,
	morestring=[b]',
	morestring=[b]"
}

\title{\epochs: Improving Blockchain Scalability\\ with Byzantine Distributed
  Sets and Barriers}

\author{
  \IEEEauthorblockN{Margarita Capretto\IEEEauthorrefmark{1},
    Martín Ceresa\IEEEauthorrefmark{1},
    Antonio Fernández Anta\IEEEauthorrefmark{2},
    Antonio Russo\IEEEauthorrefmark{2},
    César Sánchez\IEEEauthorrefmark{1}}
  \IEEEauthorblockA{\IEEEauthorrefmark{1}IMDEA Software Institute, Pozuelo de Alarcón, Madrid, Spain}
  \IEEEauthorblockA{\IEEEauthorrefmark{2}IMDEA Networks Institute, Leganés, Madrid, Spain}
  \IEEEauthorblockA{Email:\{margarita.capretto,martin.ceresa,antonio.fernandez,antonio.russo,cesar.sanchez\}@imdea.org}
    }

\date{\today}

\begin{document}

\maketitle

\begin{abstract}
  %
  %
  Blockchain technologies are facing a scalability challenge, which
  must be overcome to guarantee a wider adoption of the technology.
  This scalability issue is mostly caused by the use of consensus
  algorithms to guarantee the total order of the chain of blocks (and
  of the operations within each block).
  However, total order is often overkilling, 
  since important advanced applications of
  smart-contracts do not require a total order of
  \emph{all} the operations.
  Hence, if a more relaxed partial order (instead of a total order) 
  is allowed under certain safety conditions, a much higher scalability 
  can be achieved.
  
  %
  %
  In this paper, we propose a distributed concurrent data type, called
  \emph{\epochs}, that allows implementing this partial order and 
  increases significantly blockchain scalability.
  A \epochs implements a \emph{grow-only set object} whose elements
  are not totally ordered, unlike conventional blockchain operations.
  When convenient, the \epochs allows forcing a synchronization
  barrier that assigns permanently an epoch number to a subset of
  the latest elements added.  With the \epochs, operations in the same
  epoch are not ordered, while operations in different epochs are.
  %
  %
  We present different Byzantine-tolerant implementations of \epochs,
  prove their correctness and report on an empirical evaluation of a
  direct implementation.
  
  %
  %
  Our results show that \epochs is orders of magnitude faster than
  consensus-based ledgers to implement grow-only sets with epoch
  synchronization.
  Since the \epochs barriers can be synchronized with block
  consolidation, \epochs objects can be used as a \emph{sidechain} to
  implement many smart contract solutions with much faster operations
  than on basic blockchains.
\end{abstract}



\section{Introduction}
\label{sec:introduction}

\subsection{The Problem}

%
%
\emph{Distributed ledgers} (also known as \emph{blockchains}) were first proposed
by Nakamoto in 2009~\cite{nakamoto06bitcoin} in the implementation of Bitcoin, 
as a method to eliminate trustable third parties in electronic payment systems.
Modern blockchains incorporate smart
contracts~\cite{szabo96smart,wood2014ethereum}, which are state-full
programs stored in the blockchain that describe the functionality of
the transactions, including the exchange of cryptocurrency.
Smart contracts allow to describe sophisticated functionality, enabling
many applications in decentralized finances (DeFi)\footnote{As of December 2021, the monetary value locked in DeFi
  was estimated to be around \$100B, according to Statista
  \url{https://www.statista.com/statistics/1237821/defi-market-size-value-crypto-locked-usd/}.}, 
  decentralized governance,
Web3, etc.

The main element of all distributed ledgers is the ``blockchain,''
which is a distributed object that contains, packed
in blocks, the ordered list of transactions performed on behalf of the
users~\cite{anta2018formalizing,anta2021principles}.
This object is maintained by multiple servers without a central authority by using
consensus algorithms that are resilient to Byzantine attacks.

%
%
However, a current major obstacle for a faster widespread adoption of blockchain
technologies, and the deployment of new applications based on them, is their
limited scalability, due to the delay introduced by Byzantine
consensus
algorithms~\cite{Croman2016ScalingDecentralizedBlockchain,Tyagi@BlockchainScalabilitySol}.
Ethereum~\cite{wood2014ethereum}, one of the most popular blockchains,
is limited to less than 4 blocks per minute, each containing less than
two thousand transactions.
Bitcoin~\cite{nakamoto06bitcoin} offers even lower throughput.
These figures are orders of magnitude slower than what many
decentralized applications require, and can ultimately jeopardize the
adoption of the technology in many promising domains.
This limit in the throughput of the blockchain also increases the price per
operation, due to the high demand to execute operations.

Consequently, there is a growing interest in techniques to improve the
scalability of
blockchains~\cite{Zamani2018RapidChain,Xu2021SlimChain}.
%
Approaches include (1) the search for faster consensus
algorithms~\cite{Wang2019FastChain}, (2) the use of parallel
techniques, like sharding~\cite{Dang2019Sharding}, (3) building
application-specific blockchains with Inter-Blockchain Communication
capabilities \cite{wood2016polkadot}, \cite{kwon2019cosmos}, or (4)
extracting functionality out of the blockchain, while trying to
preserve the guarantees of the blockchain (an approach known as
``layer 2''~\cite{Jourenko2019SoKAT}).
Layer~2 (L2) approaches include the computation off-chain of Zero-Knowledge
proofs~\cite{Sasson2014ZKvonNeumann}, which only need to be checked
on-chain (hopefully more efficiently)~\cite{Sasson2014ZeroCash}, the
adoption of limited (but useful) functionality like \emph{channels}
(see, e.g., Lightning~\cite{Poon2016lightning}), or the deployment of
optimistic rollups (e.g., Arbitrum~\cite{Kalodner2018Arbitrum}) based
on avoiding running the contracts in the servers (except when needed to
annotate claims and resolve disputes).

In this paper, we propose an alternative approach to increase blockchain
scalability that exploits the following observation.
%
%
It has been traditionally assumed that cryptocurrencies require 
total order to guarantee the absence of double-spending.
However, in reality, many useful applications and functionalities 
(including cryptocurrencies~\cite{DBLP:conf/podc/GuerraouiKMPS19}) 
can tolerate more relaxed guarantees, where
operations are only \emph{partially ordered}.
%
Hence, we propose a Byzantine-fault tolerant implementation of a distributed
grow-only set, equipped with the additional operation of introducing
points of synchronization (where all servers agree on the contents of
the set).
Between barriers, elements of the distributed set can temporarily be
known by some but not all servers.
%
We call this distributed data structure a \epochs.
A blockchain \(\mathcal{B}\) implementing \epochs (as well as blocks) can align the
consolidation of the blocks of \(\mathcal{B}\)
with synchronizations, obtaining a very efficient set object as side
data type, with the same Byzantine-tolerance guarantees that \(\mathcal{B}\)
itself offers.

There are two extreme implementations of a transaction set with epochs (like \epochs)
in the context of blockchains:
\paragraph{Completely off-chain solution}
%
One could implement a transaction set totally off-chain
(either centralized or distributed), but the resulting
implementation does not have the trustability and accountability
guarantees that blockchains offer.
One example of such an attempt to implement this kind of off-chain
data types is \emph{mempools}.
Mempools (short for memory pools) are a P2P data type used by most blockchains
to maintain a set of pending transactions.
%
%
Mempools fulfill two objectives: (1) to prevent distributed
attacks to the servers that mine blocks and (2) to serve as a pool of
transaction requests from where block producers select operations.
Nowadays, mempools are receiving a lot of attention, since they suffer
from lack of accountability and are a source of
attacks~\cite{Saad2018DDoSMempool,Saad2019DDoSMempool}, including
front-running~\cite{Daian2020FlashBoys,Robinson2020DarkForest,Ferreira2021Frontrunner}.
Our proposed data structure, \epochs,
offers a much stronger accountability, because it is resilient to Byzantine
attacks and the contents of the set that \epochs maintains is public and cannot
be forged.

\paragraph{Completely on-chain solution}
One could use a smart-contract implementing the \epochs
data type.
For example, consider the following implementation (in a language
similar to Solidity), where \lstinline|add| is used to add elements,
and
\lstinline|epochinc| to increase epochs.
\begin{lstlisting}[language=Solidity,numbers=none]
  contract Epoch {
    uint public epoch = 0;
    set public the_set = emptyset;
    mapping(uint => set) public history;
    function add(elem data) public {
      the_set.add(data);
    }
    function epochinc() public {
      history[++epoch] = the_set.setminus(history);
    }
  }
\end{lstlisting}
%
\noindent Since \lstinline|epoch|,\lstinline|the_set|, and \lstinline|history|
are defined \lstinline|public| there is an implicit getter function for each of
them\footnote{In a public blockchain this function is not needed, since the set of elements
can be directly obtained from the state of the blockchain.}.
%
One problem of this implementation is that every time we add an
element, \lstinline|the_set| gets bigger, which can affect the
required cost to execute the contract.
%
A second more important problem is that adding elements is
\emph{slow}---as slow as interacting with the blockchain---while our
main goal is to provide a much faster data structure than the
blockchain.

Our approach lies in between these two extremes.
For any given blockchain \(\mathcal{B}\),
we propose an implementation of \epochs that (1) is much more
efficient than implementing and executing operations directly in
\(\mathcal{B}\);
(2) offers the same decentralized guarantees against
Byzantine attacks than \(\mathcal{B}\), and
(3) can be synchronized with the evolution of \(\mathcal{B}\), so contracts could
potentially inspect the contents of the \epochs.
In a nutshell, these goals are achieved by using faster operations for the
coordination among the servers (namely, reliable broadcast) for non-synchronized
element insertions, and use only a consensus like algorithm for epoch changes.

\subsection{Motivation}
\label{sec:motivation}
\epochs potential applications include:

\subsubsection{Mempool}
As mentioned above, user requests to execute operations/trasactions in a
blockchain are stored in a mempool before they are chosen by miners.
Once mined, the transactions executed are public, but the additional
information, including the time of insertions in the mempool, is lost.
Recording and studying the evolution of mempools would require an
additional object serving as a mempool \emph{log system}.
This storage must be fast enough to record every attempt of
interaction with the mempool without affecting the underlying
blockchain's performance, and thus, it should be much faster than the
blockchain itself.

\subsubsection{Front-running}
Mempools encode information about what it is about to happen in
blockchains, so anyone observing them can predict the next
operations to be mined, and take actions to their benefit.
\emph{Front-running} is the action of injecting transactions to be
executed before the observed transaction request.
This has been reported to be a relevant problem in decentralized
exchanges~\cite{Daian2020FlashBoys,Ferreira2021Frontrunner}.
More specifically, an operation request added to the mempool includes
the smart contract to be invoked, the maximum amount of \emph{gas}
that the user is willing to pay for the execution (\emph{gas cap}), and a \emph{fee} to
pay the miner.
Miners create blocks by choosing an attractive subset of transactions
from the mempool, attempting to maximize their profit, based on the gas cap and the
fee.
Hence, the higher the fee and the lower the gas cap, the more likely a request is
to be chosen by miners.
Users that observe the mempool can inject new operations with higher
priority (front-running) by offering a higher fee.
Detecting front-running attacks requires a bookkeeping mechanism.
As of today, once blocks are mined and consolidated, there is no
evidence of the accesses to the mempool.
The \epochs data type can serve as a basic mechanism to build a mempool
that is efficient and serves as a log of requests.

\subsubsection{Scalability by L2 Optimistic Rollups}

Optimistic rollups, like Arbitrum~\cite{Kalodner2018Arbitrum}, are
based on the idea that a set of computing entities can safely compute
outside the blockchain and post updates on the evolution of a smart
contract.
This is an optimistic strategy where each user can propose what the
next state of the contract would be.
After some time, the contract on-chain assumes that a given proposed
step is correct and executes the effects claimed.
A conflict resolution algorithm, also part of the contract on-chain, is
used to resolve disputes.
This protocol does not require the strict total order that a
blockchain guarantees, but only a record of the actions proposed.
Conflict resolutions are reduced to checking that the execution of a smart
contract would produce the claimed effect, which can be part of the
validation of the claim, and could be performed by the maintainers of
the \epochs data type.

\subsubsection{Sidechain Data}
Finally, \epochs can also be used as a general side-chain service used
to store and modify data synchronized with the blocks.
Applications that require only to update information in the storage
space of a smart contract, like digital registries, can benefit from
faster (and therefore cheaper) methods to manipulate the storage
without invoking expensive blockchain operations.

\subsection{Contributions.}

In summary, the contributions of the paper are the following:
\begin{compactitem}
\item the design and implementation of a side-chain data
  structure called \emph{distributed \epochs}.
\item several implementation of \epochs, providing different
levels of abstraction and algorithmic implementation 
improvements.
\item an empirical evaluation of a prototype implementation which
  suggests that \epochs is several orders of magnitude faster than
  consensus.
\end{compactitem}

The rest of the paper is organized as follows.
Section~\ref{sec:prelim} contains the preliminaries.
Section~\ref{sec:apisol} describes the intended properties of~\epochs.
Section~\ref{sec:implementation} describes three different
implementations of \setchain, and
Section~\ref{sec:properties} proves the correctness of the fastest
algorithm.
Section~\ref{sec:empirical} discusses an empirical evaluation of our
implementations of the different algorithms.
Section~\ref{sec:wrappers} shows how to make the use of \epochs more
robust against Byzantine servers.
Finally, Section~\ref{sec:discussion} concludes the paper.


\section{Preliminaries}\label{sec:prelim}

In this section, we present the model of computation as well as the building
blocks used in our \epochs algorithms.

\subsection{Model of Computation}

We consider a distributed system consisting of processes---clients
and servers---with an underlying communication graph in which each
process can communicate with every other process.
The computation proceeds \emph{asynchronously}, and the communication
is performed using \emph{message passing}.
Each process computes independently and at its own speed, and the
internals of each process remain unknown to other processes.
Message transfer delays are arbitrary but finite and also remain
always unknown to processes.
The intention is that servers will communicate among themselves to
implement a distributed data type with certain guarantees, and clients
can communicate with servers to exercise the data type.

Processes can fail arbitrarily, but the number of failing servers is
bounded by \(f\), and the total number of servers, \(n\), is at least
\(3f+1\).
For clients we assume that any of them can be Byzantine and we
explicitly state when we assume a client to be correct.
We assume \emph{reliable channels} between non-Byzantine (correct)
processes.
In other words, no message is lost, duplicated or modified.

Each process (client or server) has a pair of public and private
keys.
The public keys have been distributed reliably to all the processes
that may interact with each other.
Therefore, we discard the possibility of spurious or fake processes.
We assume that messages are authenticated, so that messages corrupted
or fabricated by Byzantine processes are detected and discarded by
correct processes~\cite{Cristin1996AtomicBroadcast}.
%
%
As result, communication between correct processes is reliable but
asynchronous.

Finally, we also assume that there is a mechanism for clients to create ``valid
objects'' and for servers to locally check whether an object is valid.
In the context of blockchains, using public-key crytography 
clients can sign well-formed objects and
servers can locally and efficiently check the signatures.

\subsection{Building Blocks}
We will use four building blocks in our implementations of \epochs:
Byzantine Reliable Broadcast~\cite{DBLP:journals/iandc/Bracha87,Raynal2018FaultTolerantMessagePassing},
Byzantine Atomic Broadcast~\cite{Defago2004BAB}, Byzantine
Distributed Grow-Only Sets~\cite{Cholvi2021BDSO} and Set Binary
Consensus, as described in RedBelly~\cite{Crain2021RedBelly}.
We briefly describe each separately.

\subsubsection{Byzantine Reliable Broadcast (BRB)}
The BRB service allows to broadcast
messages to a set of processes guaranteeing that messages sent by correct processes are
eventually received by \emph{all} correct processes and that all
correct processes eventually receive \emph{the same} set of messages.
The service provides a primitive \(\BRBbroadcast(m)\) for sending messages and
an event \(\BRBdeliver(m)\) for receiving messages.
%
Some important properties of BRB are:
\begin{compactitem}
\item\textbf{BRB-Validity:}\label{BRB-Validity} If a correct process
  \(p_{i}\) executes $\BRBdeliver(m)$ and $m$ belongs to a correct process
  \(p_{j}\), then \(p_{j}\) executed $\BRBbroadcast(m)$ in the past.
\item\textbf{BRB-Termination:}\label{BRB-Termination2} If a correct
  process $p$ executes $\BRBdeliver(m)$, then all correct processes
  (including $p$) eventually execute $\BRBdeliver(m)$.
\end{compactitem}
Note that BRB does not guarantee the delivery of messages in the same
order to two different correct participants.

\subsubsection{Byzantine Atomic Broadcast (BAB)}

The BAB service extends BRB with an additional
guarantee: a total order of delivery of the messages.
BAB provides the same operation and event as BRB, which we will rename
as \(\BABbroadcast(m)\) and \(\BABdeliver(m)\).
In addition to the guarantees provided by BRB services, BAB services also provide:
\begin{compactitem}
\item\textbf{Uniform Total Order:}\label{BAB-UniformTO} If processes \(p\) and \(q\)
both \(\BABdeliver(m)\) and \(\BABdeliver(m')\), then \(p\) delivers \(m\)
before \(m'\), if and only if \(q\) delivers \(m\) before \(m'\).
\end{compactitem}

%
Solving atomic broadcast has been proven to be as hard as
consensus~\cite{Defago2004BAB}, and thus, is subject to the
same limitations~\cite{Fischer1985Impossibility}.
%

\subsubsection{Byzantine Distributed Grow-only Sets (DSO)}
%
%
Sets are one of the most basic and fundamental data structures in
computer science, which typically includes operations for adding and
removing elements.
Adding and removing operations
do not commute, and thus, distributed implementations require additional
mechanisms to keep replicas synchronized to prevent
conflicting local states.
One solution is to allow only additions, forbidding removals.
%
The resulting data-type is called a grow-only set.
A grow-only set is a conflict-free replicated data
structure~\cite{Shapiro2011CCRDT} that behaves as a set in which
elements can only be added but not removed.

Let $A$ be an alphabet of values.
A grow-only set $\GS$ is a concurrent object maintaining an internal
set $\GS.S \subseteq A$ offering two operations for any process $p$:
\begin{compactitem}
  \item $\GS.\<add>(r):$ adds an element $r\in A$ to the set $\GS.S$.
  \item $\GS.\<get>():$ retrieves the internal set of elements \(\GS.S\).
\end{compactitem}
Initially, the set $\GS.S$ is empty.
A Byzantine distributed grow-only set object (DSO) is a concurrent grow-only set
implemented in a distributed manner~\cite{Cholvi2021BDSO} and tolerant to
Byzantine attacks.
%
%
Some important properties of these DSOs are:
\begin{compactitem}
\item \textbf{Byzantine Completeness}: All \(\<get>()\) and \(\<add>()\)
  operations invoked by correct clients eventually complete.
\item \textbf{DSO-AddGet}: All \(\<add>(r)\) operations will
  eventually result in $r$ being in the set returned by \emph{all} $\<get>()$. 
\item \textbf{DSO-GetAdd}: Each element $r$ returned by $\<get>()$
  was added using $\<add>(r)$ in the past.
\end{compactitem}
%


\subsubsection{Set Binary Consensus (SBC)}
SBC, introduced in
RedBelly~\cite{Crain2021RedBelly}, is a Byzantine-tolerant distributed
problem, similar to consensus.
In SBC, each participant proposes a set of elements (in the particular case of
RedBelly, a set of transactions).
After SBC finishes, all correct servers agree on a set of
valid elements which is guaranteed to be a subset of the union of
the proposed sets.
An intuitive way to understand SBC is that it efficiently 
runs binary consensus to agree on the sets proposed by each
participant, such that if the outcome is positive then the set
proposed is included in the final set consensus.
Some properties of SBC are:
\begin{compactitem}
\item \textbf{SBC-Termination}: every correct process eventually decides a
  set of elements.
\item \textbf{SBC-Agreement}: no two correct process decide different
  sets of elements.
\item \textbf{SBC-Validity}: when SBC is used on sets of transactions,
the decided set of transactions is a valid non-conflicting subset 
of the union of the proposed sets.
\item \textbf{SBC-Nontriviality}: if all processes are correct and propose
  an identical (valid non-conflicting, if transactions) set of elements, then this
  set is the decided set.
\end{compactitem}

In~\cite{Crain2021RedBelly}, the authors show that the RedBelly algorithm, 
solves SBC in a system with partial synchrony: there is an 
unknown global stabilization time after which communication is synchronous. 
Then, they propose to use SBC to replace consensus algorithms in blockchains.
They seek to improve scalability, because all transactions to be included in the
next block can be decided with one execution of the SBC algorithm.
It can be guaranteed that a block contains only valid non-conflicting
transactions by applying a deterministic function that totally orders
the decided set of transactions, and by using that order remove
invalid or conflicting transactions.

Our use of SBC is different from implementing a blockchain.
We use it to synchronize the barriers between local views of distributed
grow-only sets.
%
To guarantee that all elements are eventually assigned 
to epochs, we need the following property.
\begin{compactitem}
\item \textbf{SBC-Censorship-Resistance}: there is a time $\tau$ after which,
if the proposed sets of all correct processes 
contain the same element $e$, then $e$ will be in the decided set.
\end{compactitem}
In RedBelly, this property holds because after the global stabilization time, 
all set consensus rounds decide sets from correct processes.


\section{The \setchain Distributed Data Structure}\label{sec:apisol}

Our goal is to devise a distributed Byzantine-fault tolerant data
structure, \setchain, that implementing a grow-only set together with
synchronization barriers.
%
A key concept of Setchains is the \emph{epoch} number, which is a
global counter that the distributed data structure maintains.
The ``synchronization barrier'' is realized as an epoch change: 
the epoch number is increased and the 
elements added to the grow-only set since the previous
barrier are stamped with the new epoch number.

\subsection{API and Server State of the \setchain}

We consider a universe $E$ of elements that client processes can
inject into the set.
We also assume that servers can locally validate an element $e \in E$.
A \textbf{\setchain} with elements in \(E\) is a distributed data
structure where a set of server nodes, \(\mathbb{D}\), maintain:
\begin{compactitem}
\item a set $\<theset> \subseteq E$ of elements added;
\item a natural number $\<epoch> \in \mathbb{N}$ that denotes the latest epoch;
\item a map $\<history> : [1 \ldots \<epoch>] \rightarrow \mathcal{P}(E)$, that
  describes the sets of elements that have been stamped with an epoch number ($\mathcal{P}(E)$ denotes the power set of $E$.)
\end{compactitem}
Each server node $v \in \mathbb{D}$ supports three operations,
available to any client process:
\begin{compactitem}
\item $v.\<add>(e)$: requests that element $e$ is added to the set $\<theset>$.
\item $v.\<get>()$: returns the values of $\<theset>$, $\<history>$, and $\<epoch>$, 
as seen by $v$.
\item $v.\<epochinc>(h)$ triggers an epoch change (i.e., a
  synchronization barrier).  It must hold that $h=\<epoch>+1$.
\end{compactitem}
Informally, a client process $p$ invokes a $v.\<get>()$ operation in
node $v$ to obtain $(S,H,h)$, which is $v$'s view of set $v.\<theset>$
and map $v.\<history>$, with domain $[1\ldots h]$.
Process $p$ invokes $v.\<add>(e)$ to insert a new element $e$ in
$v.\<theset>$, and a $v.\<epochinc>(h+1)$ to request an epoch increment.
At server $v$, the set $v.\<theset>$ contains the elements that have
been added, including those that have not been
assigned an epoch yet, while \(v.\<history>\) contains only those elements that
have been assigned an epoch.
A typical scenario is that an element \(e \in E\) is first perceived
by $v$ to be in $\<theset>$, to eventually be stamped and copied to
$\<history>$ in an epoch increment.
However, as we will see, some implementations allow other ways
to insert elements, in which $v$ gets to know $e$ for the first time
during an epoch change.
The operation $\<epochinc>()$ initiates the process of collecting
elements in $\<theset>$ at each node and collaboratively decide which
ones are stamped with the current epoch.

Initially, both $\<theset>$ and $\<history>$ are empty and $\<epoch> = 0$ in
every correct server.
%
%
%
Note that client processes can insert elements to $\<theset>$ through
$\<add>()$, but only servers decide how to update $\<history>$, which
client processes can only influence by invoking $\<epochinc>()$.

%
Different servers may have, at a given point in time, different views of
the set $\<theset>$.
The \setchain data structure we propose here only provides eventual consistency
guarantees, as defined below.

\subsection{Desired Properties}
We specify now properties that a correct implementation of a
\setchain must have.
%
%
We provide here a low-level specification assuming that every 
client interaction is initiated by a
\emph{correct} client and that it interacts with a \emph{correct} server 
(recall that added elements $e$ can be locally validated
by servers).
Later, in Section~\ref{sec:wrappers}, we will describe a protocol that
allows correct clients interact with Byzantine servers, which allows
to provide a concurrent distributed object specification and
implementation, at a price in performance.

We start by requiring from a \setchain that every
$\<add>$,  $\<get>$, and $\<epochinc>$ operation issued by a 
correct client to a correct server eventually terminates.




We say that element $e$ is in epoch $h$ in history $H$ (e.g., returned by a $\<get>$
invocation) if $e\in H(h)$.
%
We say that element $e$ is in $H$ if there is an epoch $h$ such that
$e \in H(h)$.
%
%
The first property states that epochs only contain elements
coming from the grow-only set.
\newcounter{prop:consistent-set}
\setcounter{prop:consistent-set}{\value{property}}

\begin{property}[Consistent Sets]\label{api:consistent-set}
  Let $(S,H,h)=v.\<get>()$ be the result of an invocation to a correct server $v$.
  Then, for each $i\leq h, H(i) \subseteq S$.
\end{property}
The second property states that every element added to a correct server is
eventually returned in all future gets issued on the same server.
\begin{property}[Add-Get-Local]\label{api:history->theset-local}
  Let  $v.\<add>(e)$ be an operation invoked by a correct client 
  to a correct server $v$.
  Then, eventually all invocations $(S,H,h)=v.\<get>()$ 
  satisfy $e\in S$.
\end{property}
The next property states that elements present
in a correct server are propagated to all correct servers.
\begin{property}[Add-Get]\label{api:history->theset}
  Let $v,w$ be two correct servers, let $e \in E$ and let
  $(S,H,h)=v.\<get>()$.
  If $e \in S$, then eventually all invocations
  $(S',H',h')=w.\<get>()$ satisfy that $e \in S'$.
\end{property}
We assume in the rest of the paper that at every point in time, there is a
future instant at which $\<epochinc>()$ is invoked and completed.
This is a reasonable assumption in any real practical scenario, since
it can be easily guaranteed using timeouts.
Then, the following property states that all elements added are eventually
assigned an epoch.
\begin{property}[Eventual-Get]\label{api:theset->history}
  Let $v$ be a correct server, let $e \in E$ and let $(S,H,h)=v.\<get>()$.
  If $e \in S$, then eventually all invocations
  $(S',H',h')=v.\<get>()$ satisfy that $e \in H'$.
\end{property}
The previous three properties imply the following property.
\begin{property}[Get-After-Add]\label{api:get-after-add}
  Let $v.\<add>(e)$ be an operation invoked by a correct client to
  a correct server $v$ and valid element $e \in E$.
  Then, eventually all invocations $(S,H,h)=w.\<get>()$ 
  satisfy that $e\in H$, for all correct servers $w$.
\end{property}
%
%
An element can be at most in one epoch.
\begin{property}[Unique Epoch]\label{api:local_unique_stamp}
  Let $v$ be a correct server,
  $(S,H,h)=v.\<get>()$, and let
  $i,i'\leq{}h$ with $i\neq i'$.
  Then, $H(i)\cap{}H(i')=\emptyset$.
\end{property}
All correct server processes agree on the content of every epoch (a safety
property).
\begin{property}[Consistent Gets]\label{api:consistent-gets}
  Let $v,w$ be correct servers, let $(S,H,h)=v.\<get>()$ and
  $(S',H',h')=w.\<get>()$, and let $i\leq \min(h,h')$. Then
  $H(i)=H'(i)$.
\end{property}
The two previous properties imply that no element can be in two
different epochs even if the history sets are obtained from $\<get>$
invocations to two different (correct) servers.
%
%
%
Then, Property~\ref{api:consistent-gets} essentially states that the
histories returned by two $\<get>$ invocations to correct servers 
are one the prefix of the other.
For comparison, it is not necessarily true for the $\<theset>$ part of
$\<get>$ that two invocations return sets that are contained one in
the other.
The reason is that the \setchain data structure allows fast insertion
of elements that takes time to propagate to all correct nodes, so if
two elements $e$ and $e'$ are inserted at two different correct
servers, some correct servers may know $e$ but not $e'$, and vice
versa.

Finally we require that every element in the history comes from the
result of a client adding the element.
%
%
%
\begin{property}[Add-before-Get]\label{api:get->add}
  Let $v$ be a correct server, $(S,H,h)=v.\<get>()$,
  and $e \in S$.
  Then, there was an operation $w.\<add>(e)$ in the past.
\end{property}


\section{Implementations}\label{sec:implementation}

In this section, we describe implementations of the \setchain data
structure that satisfy the properties in Section~\ref{sec:apisol}.
We first describe a centralized sequential implementation, and then
three distributed implementations.
The first distributed implementation is built using a Byzantine 
distributed grow-only set
object (DSO) to maintain $\<theset>$, 
and Byzantine atomic broadcast (BAB) for epoch 
increments.
%
The second distributed implementation is also built using DSO, but it
uses Byzantine reliable broadcast (BRB) to announce epoch increments
and set binary consensus (SBC) for epoch changes.
Finally, the third implementation uses local sets, BRB for
broadcasting elements and epoch increment announcements, and SBC for
epoch changes.

\subsection{Sequential Implementation}

\addtocounter{algorithm}{-1} 
\begin{figure}[b!]
    \begin{adjustbox}{minipage=[t]{\columnwidth}}
      \begin{algorithm}[H]
        \caption{\small Single server implementation.}%
              \label{alg:sequential}%
              \begin{algorithmic}[1]
                \State \textbf{Init:} $\<epoch> \leftarrow 0,$
                \hspace{2em}
                $\<history> \leftarrow \emptyset$\label{seq:history} 
                \State \textbf{Init:} $\<theset> \leftarrow \emptyset$\label{seq:theset}
                \Function{\<Get>}{~}
                  \State \textbf{return} $(\<theset>, \<history>, \<epoch>)$
                \EndFunction
                \Function{\<Add>}{$e$}
                  \State \textbf{assert} {\(valid(e)\)}
                  \State \(\<theset> <- \<theset> \cup \{e\}\)
                \EndFunction
                \Function{\<EpochInc>}{$h$}
                    \State \textbf{assert} $h \equiv \<epoch> + 1$
                    \State \(\textit{proposal} <- \<theset> \setminus \bigcup_{k=1}^{\<epoch>} \<history>(k)\)
                    \State $\<history> \leftarrow \<history> \cup \{\langle h, \textit{proposal} \rangle\}$
                    \State $\<epoch> \leftarrow \<epoch> + 1$
                \EndFunction
              \end{algorithmic}
            \end{algorithm}
      \end{adjustbox}
  \end{figure}

Alg.~\ref{alg:sequential} shows a centralized solution, which
maintains two local sets, $\<theset>$ and $\<history>$, both
initialized as empty sets.
The set $\<theset>$ records all added elements and $\<history>$ is
implemented as a collection of pairs $\langle h,A \rangle$ where $h$ is an epoch
number and $A$ is a set of elements. We use $\<history>(h)$ to refer to the
set $A$ in the pair $\langle h,A \rangle \in \<history>$.
In this implementation, it is easy to maintain a local copy for
$\<theset>$ because there is a single node maintaining the \setchain.
Additionally, the implementation keeps a natural number $\<epoch>$
that is incremented each time there is a new epoch.
The implementation of the data structure is as follows:
$\<Add>(e)$
checks that element $e$ is valid and adds it to $\<theset>$.
$\<Get>()$ returns $(\<theset>,\<history>,\<epoch>)$. 
Returning the latest epoch number, $\<epoch>$, allows
clients invoke $\<EpochInc>(h)$ with $h=\<epoch>+1$.

\begin{figure}[t!]
  \begin{adjustbox}{minipage=[t]{\columnwidth}}
    \begin{algorithm}[H]
      \caption{\small Server $i$ implementation using DSO and BAB}%
      \label{DPO-alg1}%
      \begin{algorithmic}[1]
                \State \textbf{Init:} $\<epoch> \leftarrow 0,\hspace{2em}\<history>\leftarrow\emptyset$\label{alg1:history}
            \State \textbf{Init:} $\<theset> \leftarrow $\<DSO>.\<Init>\(()\)~\label{alg1:init_e}
            \Function{\<Get>}{~}
              \State \<return> $(\<theset>.\text{\<Get>}(), \<history>,\<epoch>)$~\label{alg1:func_get}
            \EndFunction
            \Function{\<Add>}{$e$}
            \State \textbf{assert} {\(valid(e)\)}
            \State $\<theset>.{\<Add>}(e)$
            \EndFunction
            \setcounter{ALG@line}{11}
            \Function{\<EpochInc>}{$h$}\label{alg1:epoch_inc}
              \State \textbf{assert} $h \equiv \<epoch> + 1$
              \State $proposal \leftarrow
                \<theset>.\text{\<Get>}()
                \setminus \bigcup_{k=1}^{\<epoch>} \<history>(k)$~\label{alg1:proposal}
              \State \<BAB>.\<Broadcast>(\<mepochinc>($h, proposal, i$))\label{alg1:bcast_proposal}
            \EndFunction
            \Upon{\<BAB>.\<Deliver>(\<mepochinc>($h, proposal,
                  j$))\\\hspace{2em} from $2f+1$ different servers $j$ for the same $h$}\label{alg1:upon_delivery} 
              \State \textbf{assert} $h \equiv \<epoch> + 1$
              \State $E \leftarrow \{e:e \in \textit{proposal}$ \\
              \hspace{4em}for at least $f+1$ different $(h,proposal,j)\}$
              \State $\<history> \leftarrow \<history> \cup \{\langle h, E \rangle\}$\label{alg1:added}
              \State $\<epoch> \leftarrow \<epoch> + 1$
            \EndUpon
          \end{algorithmic}
        \end{algorithm}
        \end{adjustbox}
  \end{figure}

\subsection{Distributed Implementations}

We present three distributed algorithms beginning with a data
structure that uses off-the-self existing building blocks.
We then present more efficient implementations.

\subsubsection{First approach. DSO and BAB}
Alg.~\ref{DPO-alg1} uses two external services: a DSO and BAB.
%
We denote messages with the name of the message followed by its
content as in ``$\<mepochinc>(h,\textit{proposal},i)$''.
The variable $\<theset>$ is not a local set anymore, but a DSO initialized
empty with $\<Init>()$ in line~\ref{alg1:init_e}.
The function $\<Get>()$ invokes the DSO $\<Get>()$ function
(line~\ref{alg1:func_get}) to fetch the set of elements.
The function $\<EpochInc>(h)$ triggers the mechanism required to
increment an epoch and reach a consensus on which elements should
be in it.
%
This process begins by computing a local $\textit{proposal}$ set, of
those elements added but that have not been stamped with an epoch
yet (line~\ref{alg1:proposal}).
The $\textit{proposal}$ set is then broadcasted using a BAB service
alongside the epoch number $h$ and the server node id $i$
(line~\ref{alg1:bcast_proposal}).
Then, the server waits to receive exactly $2f+1$ proposals, and keeps
the set of elements $E$ present in at least $f+1$ proposals, which
guarantees that each element $e \in E$ was proposed by at least one correct
server.
The use of BAB guarantees that every message sent by a correct server
eventually reaches every other correct server in the \emph{same order}, so
all correct servers use the same set of $2f+1$ proposals.
Therefore, all correct servers arrive to the same conclusion, 
and the set $E$
is added as epoch $h$ in $\<history>$ in line~\ref{alg1:added}.

Alg.~\ref{DPO-alg1}, while easy to understand and prove correct, is not
efficient.
To start, in order to complete an epoch increment, it requires at
least $3f+1$ calls to $\<EpochInc>(h)$ to different servers, so at least
$2f+1$ proposals are received (the $f$ Byzantine severs may not
propose anything).
Another source of inefficiency comes from the use of off-the-shelf
building blocks.
For instance, every time a DSO $\<Get>()$ is invoked, many messages are
exchanged to compute a reliable local view of the set~\cite{Cholvi2021BDSO}.
Similarly, every epoch change requires a DSO $\<Get>()$ in
line~\ref{alg1:proposal} to create a proposal.
Additionally, line~\ref{alg1:upon_delivery} requires waiting for
$2f+1$ atomic broadcast deliveries to take place.
The most natural implementations of BAB services solve one
consensus per message delivered (see Fig.~7 in
\cite{Chandra1998FailureBAB}), which would make this algorithm very
slow.
We solve these problems in two alternative algorithms.
\begin{figure}[t!]
      \begin{adjustbox}{minipage=[t]{\columnwidth}}
  \begin{algorithm}[H]
    \caption{\small Server $i$ implementation using DSO, and reliably
      broadcast (BRB) and set consensus (SBC) (Red Belly primitives). }
    \label{DPO-alg-basb}
    \begin{algorithmic}[1]
      \setcounter{ALG@line}{10} 
      \State \ldots     \Comment{$\<Get>$ and $\<Add>$ as in Alg.~\ref{DPO-alg1}}

      \Function{\<EpochInc>}{$h$}\label{alg2:epochinc}
        \State \<assert> $h==\EPOCH +1$
        \State \<BRB>.\<Broadcast>(\<mepochinc>($h$))\label{alg2:brb-epochinc}
      \EndFunction
      \Upon{\<BRB>.\<Deliver>(\<mepochinc>($h$)) and $h<\EPOCH+1$}
        \State \<drop>
      \EndUpon
      \Upon{\<BRB>.\<Deliver>($h$) and $h==\EPOCH+1$}
        \State \<assert> $prop[h]==null$
        \State $prop[h] \leftarrow \THESET.\text{\<Get>}() \setminus 
        \bigcup_{k=1}^{\<epoch>} \<history>(k)$\label{DPO-alg2-dsoget}
        \State \<SBC>[$h$].\<Propose>($prop[h]$)
      \EndUpon
      %
      %
      \Upon{\<SBC>[$h$].\<SetDeliver>($propset$) \\ \hspace{2em}  and  $h==\EPOCH+1$} 
        \State $E \leftarrow \{e:e \in $ at least $f+1$ different $propset[j]\}$
        \State $\HISTORY \leftarrow \HISTORY \cup \{\langle h, E \rangle\}$
        \State{$\EPOCH \leftarrow \EPOCH+1$} 
        \tikz{\node(tmp2){};}
        \EndUpon
      \end{algorithmic}
    \end{algorithm}
        \end{adjustbox}
\end{figure}

\begin{figure}[t!]
	\begin{algorithm}[H]
          \caption{\small Server implementation using a local set,
            Byzantine reliable broadcast (BRB) and Byzanting set
            consensus (SBC), (Red Belly
            primitives).}~\label{DPO-alg-basb3}
		\begin{algorithmic}[1]
			\State  \textbf{Init:} $\EPOCH \leftarrow 0,$
			\hspace{2em} $\HISTORY \leftarrow \emptyset$
			\State  \textbf{Init:} $\THESET \leftarrow \emptyset$\label{alg3:theset}
			\Function{\<Get>}{~}
			        \State \<return> $(\THESET, \HISTORY,\<epoch>)$\label{alg3:get}
			\EndFunction
			\Function{\<Add>}{$e$}
                                \State \<assert> $valid(e)$ and $e \notin \THESET$
                                \State \<BRB>.\<Broadcast>(\<madd>(e)) \label{alg3:brb-broadcast}
                                \EndFunction
			\Upon{\<BRB>.\<Deliver>(\<madd>\((e)\))}
                                \State \<assert> $valid(e)$
			        \State $\THESET \leftarrow \THESET \cup \{e\}$ \label{add_theset_BRBDeliver}
			\EndUpon
			\Function{\<EpochInc>}{$h$}
                                \State \<assert> $h==\EPOCH +1$
			        \State \<BRB>.\<Broadcast>(\<mepochinc>($h$))
                        \EndFunction
			\Upon{\<BRB>.\<Deliver>(\<mepochinc>($h$)) and $h<\EPOCH+1$}
                                \State \<drop>
                        \EndUpon
			\Upon{\<BRB>.\<Deliver>(\<mepochinc>($h$)) and $h==\EPOCH+1$}
                                \State \<assert> \(prop[h] == \emptyset\)
                                \State $prop[h] \leftarrow \THESET \setminus 
                                \bigcup_{k=1}^{\<epoch>} \<history>(k)$
                                \State \<SBC>[$h$].\<Propose>($prop[h]$)
			\EndUpon
			\Upon{\<SBC>[\(h\)].\<SetDeliver>($propset$) and  $h==\EPOCH+1$}
			        \State $E \leftarrow \{e : e \in propset[j], valid(e) \wedge e\notin \HISTORY\}$~\label{forall_e_SBCDeliver}
				    \State $\HISTORY \leftarrow \HISTORY \cup \{\langle h, E \rangle\}$~\label{add_history_SBCSetDeliver}
				    \State $\THESET \leftarrow \THESET \cup E$~\label{add_theset_SBCSetDeliver}
				    \State $\EPOCH \leftarrow \EPOCH+1$\label{set_epoch_SBCSetDeliver} 
			\EndUpon
		\end{algorithmic}
	\end{algorithm}\vspace{-2em}
\end{figure}

\subsubsection{Second approach. Avoiding BAB}

Alg.~\ref{DPO-alg-basb} improves the performance of
Alg.~\ref{DPO-alg1} in several ways. First, it uses BRB to propagate
epoch increments, so a client does not need to contact more than one
server.
Second, the use of BAB and the wait for the arrival of
$2f+1$ messages in line~\ref{alg1:upon_delivery} of
Alg.~\ref{DPO-alg1} is replaced by using a SBC
algorithm, which allows solving several consensus instances
simultaneously.

On a more abstract level, what we want to happen when an $\<EpochInc>(h)$ is
triggered is that new elements in the local set $\<theset>$ of each correct node
are stamped with the new epoch number and added to the set $\<history>$.
The DSO guarantees eventual consistency, but at a given moment in
time, two invocations to $\<Get>$ may vary from server to server.
However, we need to guarantee that for every epoch the set
$\<history>$ is the same in every correct server.
Alg~\ref{DPO-alg1} enforced this using BAB and counting enough
received messages to guarantee that every stamped element $e$ was sent
by a correct node.
Alg.~\ref{DPO-alg-basb} uses SBC to solve several independent
consensus instances simultaneously, one on each participant's
proposal.
%
Line~\ref{alg2:brb-epochinc} broadcasts a message inviting servers to
begin an epoch change.
When a correct server receives an epoch change invitation, it builds a
proposed set and sends this proposal to the SBC.
Note that
there is one instance of SBC per epoch change, identified by $h$.
Finally, with SBC each correct server receives the same set
of proposals (where each proposal is a set of elements).
Then, every node applies the same function to the same set of
proposals reaching the same conclusion on how to update $\<history>(h)$.
The function applied is deterministic: preserving only elements $e$
that are present in at least $f+1$ proposed sets, so these elements
are guaranteed to have been proposed by a correct server.
Observe that Alg.~\ref{DPO-alg-basb} still requires 
several invocations of the DSO $\<Get>$ operation to build the local
proposal, one at each server.


\subsubsection{Final approach. BRB and SBC without DSOs}

The last approach, shown in Alg.~\ref{DPO-alg-basb3}, avoids the cascade
of messages that DSO $\<Get>$ calls require in Alg.~\ref{DPO-alg1} by
dissecting the internals of the DSO, and incorporating the corresponding steps in the
\setchain algorithm directly.
This idea exploits the fact that \emph{a correct} \setchain \emph{server} is a
\emph{correct client} of the DSO, and there is no need for the DSO to be defensive
(this illustrates that using Byzantine resilient building blocks does not
compose efficiently, but exploring this general idea is out of the scope of this
paper).

Alg.~\ref{DPO-alg-basb3} implements $\<theset>$ using a local set
(line~\ref{alg3:theset}) to accumulate a local view of the elements
added.
Elements newly received (in $\<Add>(e)$) will be communicated to other
server nodes using BRB.
At any given point in time two correct servers may have a different
local set (due to pending BRB deliveries) but each element added in
one will eventually be known to the other.
The local variable $\<history>$ is only updated in
line~\ref{add_history_SBCSetDeliver} with the result of a SBC round.
Therefore, all correct servers will agree on the same sets, with all
elements being valid, not previously stamped and proposed by some
server.
Valid elements, by assumption, have been provided by correct clients.
Additionally, Alg.~\ref{DPO-alg-basb3} updates $\<theset>$ to
account for elements that are new to the node, guaranteeing that all
elements in $\<history>$ are also in $\<theset>$.
Note that this opens the opportunity to add elements to the \setchain by
proposing them during an epoch change without broadcasting them before.
This is exploited in Section~\ref{sec:empirical} to speed up the
algorithm even more.
As a final note, Alg.~\ref{DPO-alg-basb3} allows a Byzantine server to 
propose elements, which will
be accepted as long as they are valid. This is equivalent to a correct
client proposing an element through an $\<Add>()$ operation, which 
is then successfully propagated during the set consensus phase.


\section{Proof of Correctness}
\label{sec:properties}

We prove now the correctness of Alg.~\ref{DPO-alg-basb3}.
%
%
We first show that all stamped elements are in $\<theset>$.

\begin{lemma}
  \label{lem:consistent-set}
  For every correct server $v$, at the end of each function/upon,
  $\bigcup_{h} v.\<history>(h) \subseteq v.\<theset>$.
\end{lemma}

\begin{IEEEproof}
  Let $v$ be a server.
  The only way to add elements to $v.\<history>$ is at
  line~\ref{add_history_SBCSetDeliver}, which is followed by
  line~\ref{add_theset_SBCSetDeliver} which adds the same elements to
  $v.\<theset>$.
  The only other instruction that modifies $v.\<theset>$ is
  line~\ref{add_theset_BRBDeliver} which only makes the set grow.
\end{IEEEproof}
Lemma~\ref{lem:consistent-set} directly implies that
Alg.~\ref{DPO-alg-basb3} satisfies Prop.~\ref{api:consistent-set}
(\textit{Consistent Sets}) from Section~\ref{sec:apisol}.


\begin{lemma}\label{lem:eventually-theset}
  Let $v$ be a correct server and $e$ an element in $v.\<theset>$.
  Then $e$ will eventually  be in $w.\<theset>$ for every correct server
  $w$.
\end{lemma}
\begin{IEEEproof}
  Initially, $v.\<theset>$ is empty. There are two ways to add an
  element $e$ to $v.\<theset>$:
  (1) At line~\ref{add_theset_BRBDeliver}, so $e$ is valid and was received via a
    $\<BRB>.\<Deliver>(\<madd>(e))$.
    By Properties \textbf{BRB-Validity} and \textbf{BRB-Termination}
    of BRB (see Section~\ref{sec:prelim}), every correct server $w$
    will eventually execute $\<BRB>.\<Deliver>(\<madd>(e))$, and then (since
    $e$ is valid), $w$ will add it to $w.\<theset>$ in
    line~\ref{add_theset_BRBDeliver}.
  (2) At line~\ref{add_theset_SBCSetDeliver}, so element $e$ is valid and was
    received as an element in one of the sets in $\Propset$ from
    $\<SBC>[h].\<SetDeliver>(\Propset)$ with $h = v.\<epoch> + 1$.
    By properties \textbf{SBC-Termination} \textbf{SBC-Agreement} and
    \textbf{SBC-Validity} of SBC (see Section~\ref{sec:prelim}), all
    correct servers agree on the same set of proposals.
    Therefore, if $v$ adds $e$ then $w$ either adds it or has it
    already in its $w.\<history>$ which implies by
    Lemma~\ref{lem:consistent-set} that $e\in w.\<theset>$.
    In either case, $e$ will eventually be in $w.\<theset>$.
\end{IEEEproof}

Lemma~\ref{lem:eventually-theset}, and the code of function $\<Add>()$ and line~\ref{alg3:get} of
function $\<Get>()$ in Alg.~\ref{DPO-alg-basb3} imply
Prop.~\ref{api:history->theset-local} (\textit{Add-Get-Local}) and
Prop.~\ref{api:history->theset} (\textit{Add-Get}) in
Section~\ref{sec:apisol}.
The following lemmas reason about how elements are stamped.

\begin{lemma}\label{lem:stamp-once}
  Let $v$ be a correct server and $e\in v.\<history>(h)$ for some
  $h$. Then, for any \(h' \neq h\),
  $e\notin v.\<history>(h')$.
\end{lemma}

\begin{IEEEproof}
  It follows directly from the check that $e$ is not injected at
  $v.\<history>(h)$ if $e\in v.\<history>$ in
  line~\ref{add_history_SBCSetDeliver}.
\end{IEEEproof}

\begin{lemma}\label{lem:history-unique}
  Let $v$ and $w$ be correct servers.
  At a point in time, let $h$ be such that $v.\<epoch> \geq h$ and
  $w.\<epoch> \geq h$. Then $v.\<history>(h)=w.\<history>(h)$.
\end{lemma}

\begin{IEEEproof}
  The proof proceeds by induction on $\<epoch>$. 
  The base case is $\<epoch>=0$, which holds trivially since
  $v.\<history>(0)=w.\<history>(0)=\emptyset$.
  Variable $\<epoch>$ is only
  incremented in one unit in line~\ref{set_epoch_SBCSetDeliver},
  after $\<history>(h)$ has been changed in
  line~\ref{add_history_SBCSetDeliver} when $h=\<epoch>+1$.
  In that line, $v$ and $w$ are in the same phase 
  on SBC (for the same $h$).
  By \textbf{SBC-Agreement}, $v$ and $w$ receive the same \Propset,
  both $v$ and $w$ validate all elements equally, and (by inductive
  hypothesis), for each $h' \leq \<epoch>$ it holds that 
  $e\in v.\<history>(h')$ if and only if $e\in w.\<history>(h')$.
  Therefore, in line~\ref{add_history_SBCSetDeliver} both $v$ and $w$
  update $\<history>(h)$ equally, and after line~\ref{set_epoch_SBCSetDeliver}
  it holds that $v.\<history>(\<epoch>)=w.\<history>(\<epoch>)$.
\end{IEEEproof}

\begin{lemma}\label{lem:eventually-history}
  Let $v$ and $w$ be correct servers. If $e\in v.\<theset>$. Then,
  eventually $e$ is in $w.\<history>$.
\end{lemma}

\begin{IEEEproof}
  By Lemma~\ref{lem:eventually-theset} every correct server $w$ will
  satisfy $e\in w.\<theset>$ at some $t > \tau$.
  By assumption, there is a new $\<EpochInc>()$ after $t$ (let the
  epoch number be $h$).
  If $e$ is already in $\<history>(h')$ for $h'<h$ we are done, since
  from Lemma~\ref{lem:history-unique}
  in this case at the end of the SBC phase for $h'$ every
  correct server node $w$ has $e$ in $w.\<history>(h')$.
  If $e$ is not in $\<history>$ at $t$ then,
  \textbf{SBC-Censorship-Resistance} guarantees that 
  the decided set will contain $e$.
  %
  Therefore, at line~\ref{add_history_SBCSetDeliver} every  correct
  server $w$ will add $e$ to $w.\<history>(h)$.
\end{IEEEproof}

Lemmas~\ref{lem:history-unique} and \ref{lem:eventually-history}
imply  that all elements will be stamped, that is,
Prop.~\ref{api:history->theset} (\textit{Eventual-Get}).
Prop.~\ref{api:get-after-add} follows from
Prop.~\ref{api:history->theset}.
Lemma~\ref{lem:stamp-once} directly implies
Prop.~\ref{api:local_unique_stamp} (\textit{Unique Epoch}).
Finally, Lemma~\ref{lem:history-unique} is equivalent to
Prop.~\ref{api:consistent-gets} (\textit{Consistent Gets}) from
Section~\ref{sec:apisol}.

Finally, we discuss Prop.~\ref{api:get->add} (\textit{Add-before-Get}).
If valid elements can only be created by correct clients, and correct
clients only interact with server nodes using the API, then the only
way to reveal $e$ to server nodes is through $\<Add>(e)$ and therefore
the property trivially holds.
If, on the other hand, valid elements can be created by, for example
Byzantine server nodes, then server nodes can inject elements in
$\<theset>$ and $\<history>$ of correct servers without using the $\<Add>$
function. They can either execut directly a BRB.$\<Broadcast>$,
 like the one in line~\ref{alg3:brb-broadcast}, 
or even directly injecting them via the SBC.
In these cases, Alg.~\ref{DPO-alg-basb3} satisfies a weaker version of
(\textit{Add-before-Get}) that states that elements returned by
$\<Get>()$ are valid and are either added by $\<Add>()$, 
by a BRB.$\<Broadcast>$ like that of line~\ref{alg3:brb-broadcast}, 
or injected in the
SBC phase.
In fact, we will use this idea to propagate elements directly in the
SBC phase to accelerate Alg.~\ref{DPO-alg-basb3}.

\newcommand{\AlgTwo}{Alg.~\ref{DPO-alg-basb}\xspace}
\newcommand{\AlgThree}{Alg.~\ref{DPO-alg-basb3}\xspace}
\newcommand{\AlgTwoSet}{Alg.~\ref{DPO-alg-basb}+set\xspace}
\newcommand{\AlgThreeSet}{Alg.~\ref{DPO-alg-basb3}+set\xspace}

\section{Empirical Evaluation}
\label{sec:empirical}

\newcommand{\Exp}[1]{\includegraphics[width=0.480\textwidth]{experiments/#1.pdf}}

\begin{figure*}[t!]
  \footnotesize\centering
  \noindent
  \begin{tabular}{@{}c@{}c@{}}
    \Exp{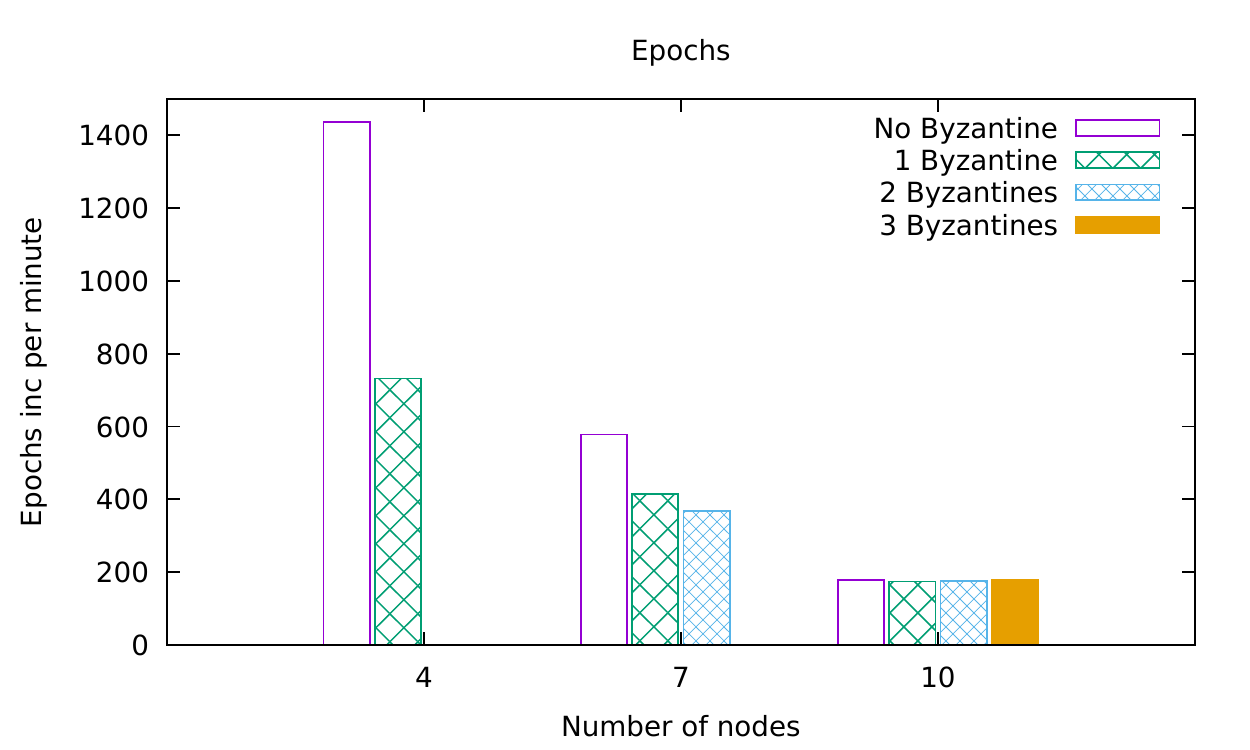} & \Exp{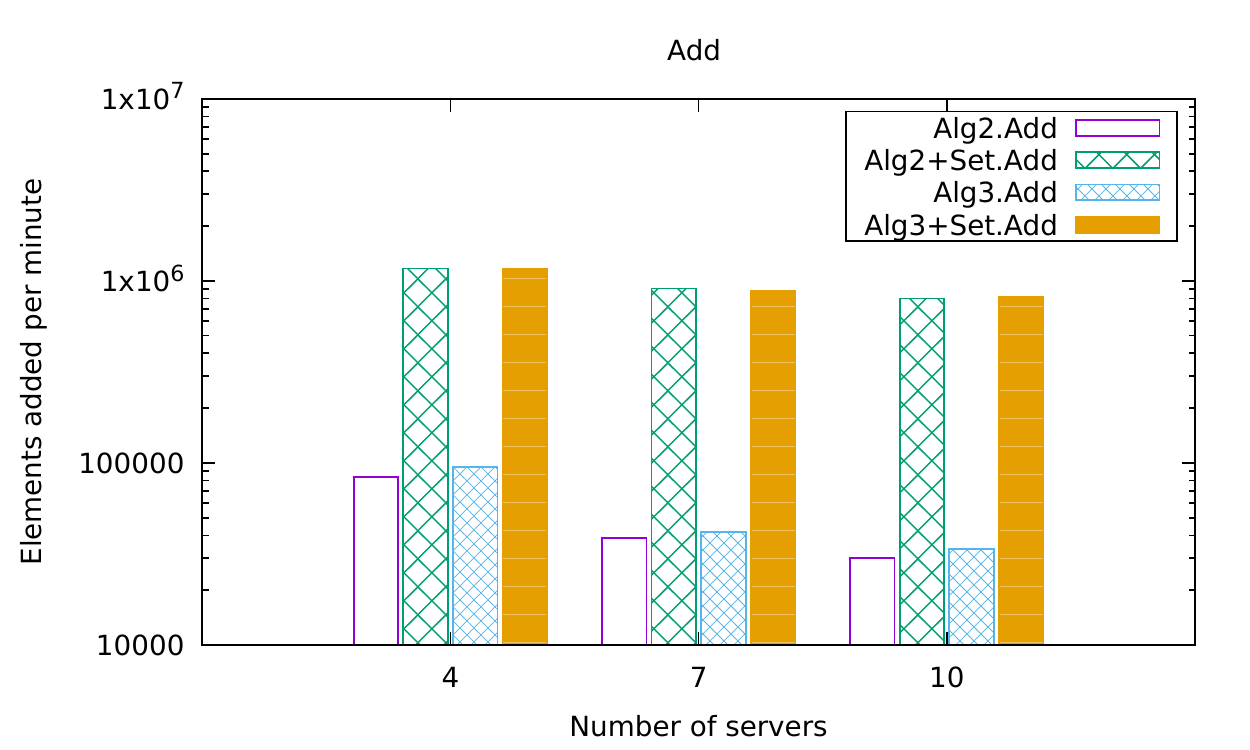} \\
    (a) Maximum epoch changes  & (b) Maximum elements added (no epochs)\\
    \Exp{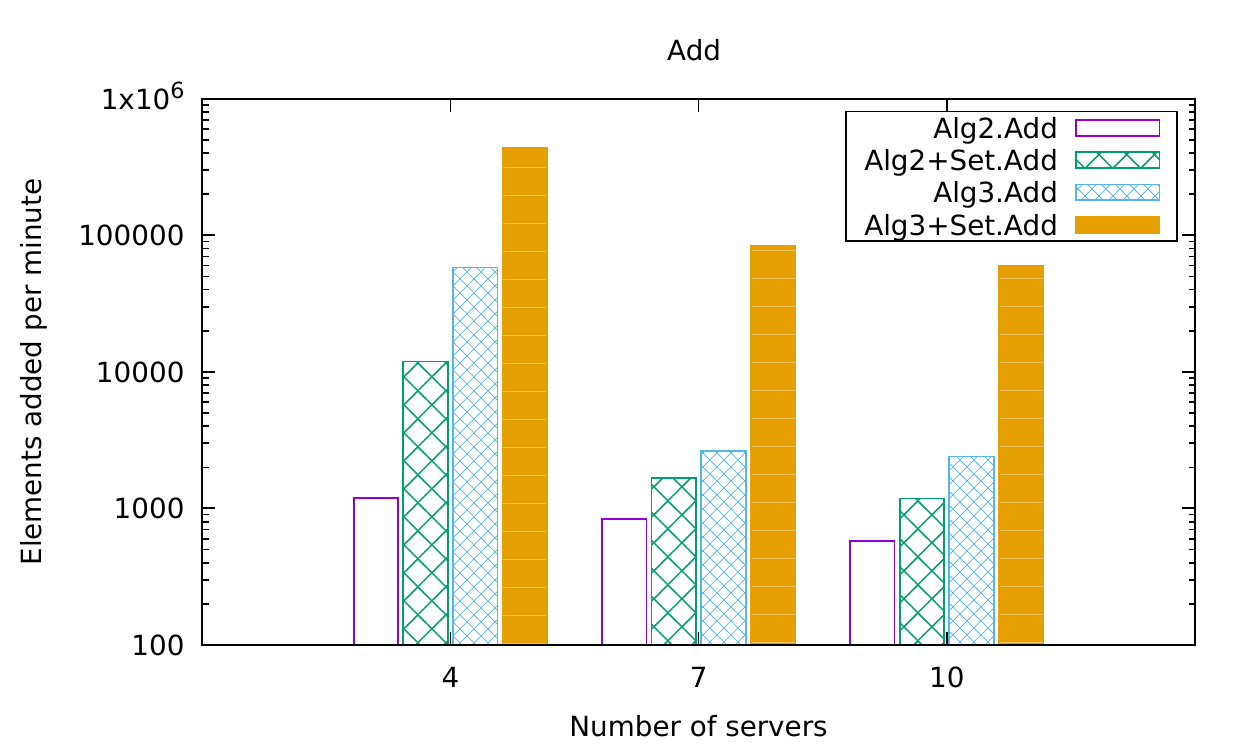} & \Exp{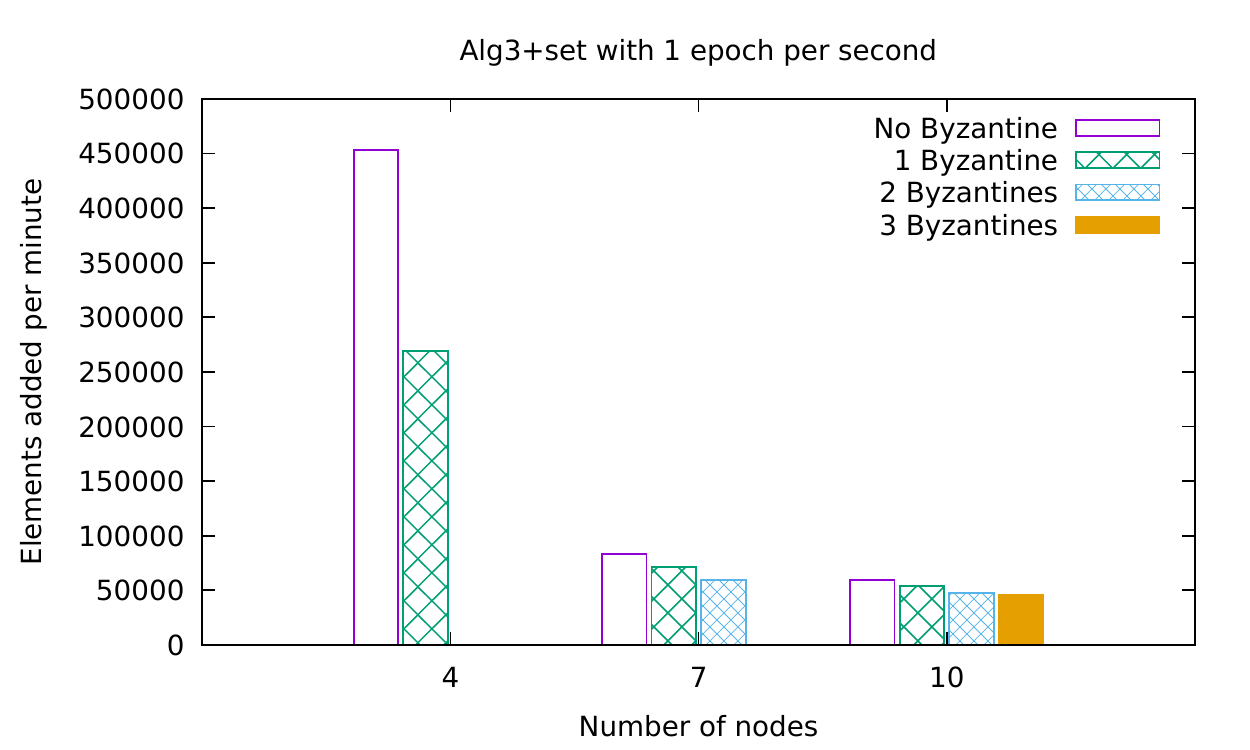} \\
    (c) Maximum elements added (with epochs) & (d) Silent Byzantine effect in adds\\
    \Exp{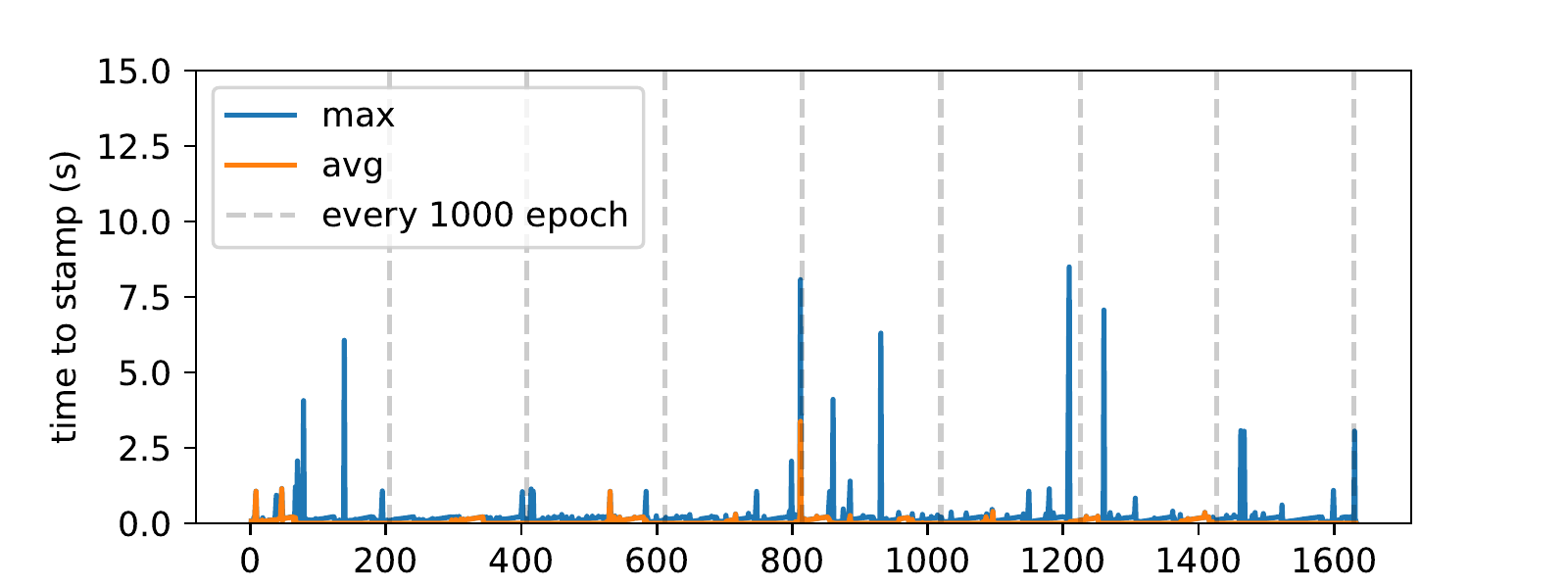} &     \Exp{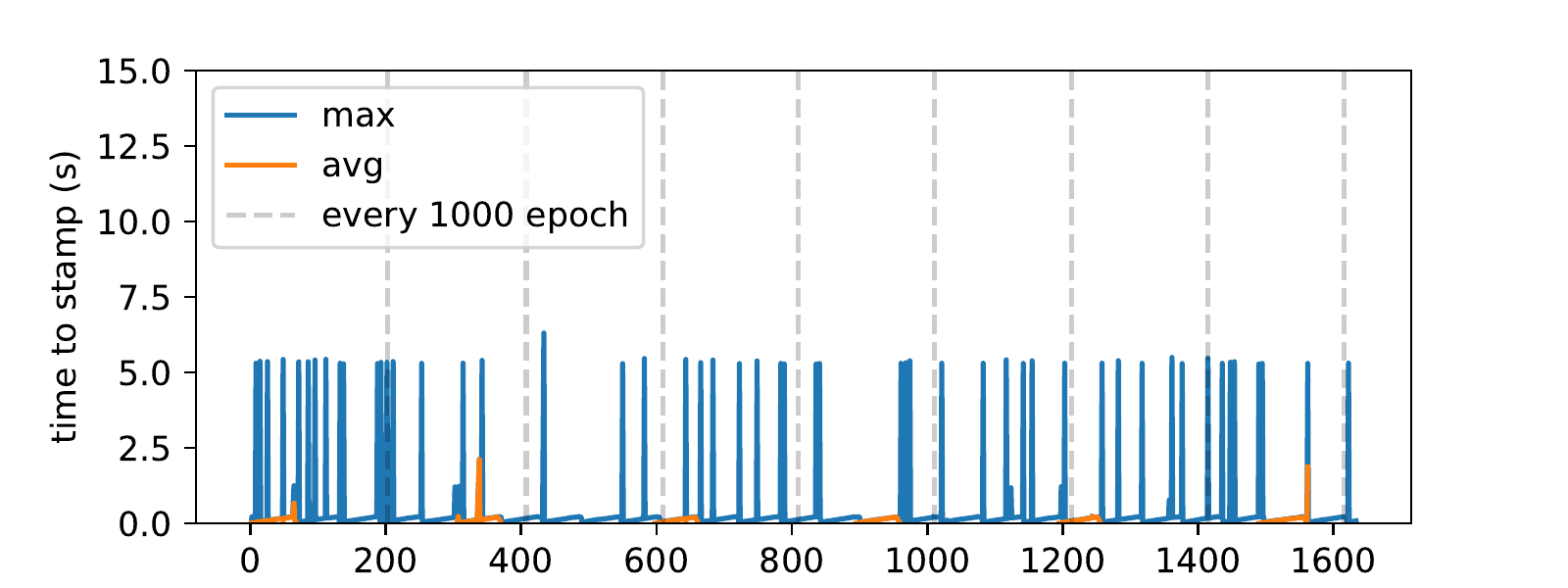} \\
    (e) Time to stamp (\AlgThree) & (f) Time to stamp (\AlgThree with aggregation)\\
  \end{tabular}
  \caption{Experimental results. \AlgTwoSet and \AlgThreeSet are the versions of the algorithms with aggregation. Byzantine servers are simply silent.}
  \label{fig:experiments}
\end{figure*}

%
%
%
We have implemented the necessary building blocks (1) DSO server code,
(2) Reliable Broadcast and (3) Set Binary Consensus servers, and
using these building blocks we have implemented \AlgTwo and \AlgThree.
%
%
Our prototype is written in Golang~\cite{donovan15go} 1.16 with
message passing using ZeroMQ~\cite{zeromq} over TCP.
Our testing platform uses Docker running on a server with 2 Intel
Xeon CPU processors at 3GHz with 36 cores and 256GB RAM, running
Ubuntu 18.04 Linux64.
Each \setchain server node was wrapped in a Docker container with no
limit on CPU or RAM usage.
\AlgTwo implements a \setchain and a DSO as two
standalone executables that communicate using remote procedure calls
on the internal loopback network interface of the Docker container.
The RPC server and client are taken from the Golang standard
library.
For \AlgThree everything resides in a single
executable.
For both algorithms, we evaluate two versions, one where each
  element inserted causes a broadcast and another where servers
  aggregate locally inserted elements until a maximum message size (of $10^6$ elements) or a
  maximum element timeout (of 5s) is reached.
  
%
%
We evaluate empirically the following hypothesis:
\begin{compactitem}
\item (\Hspeed): The maximum rate of elements that can be inserted is
  much higher than the maximum epoch rate.
\item (\Hbetteralg): Alg.~\ref{DPO-alg-basb3} performs better than
  Alg.~\ref{DPO-alg-basb}.
\item (\Hagr): The aggregated versions perform better than
  the basic versions.
\item (\Hbyzantine) Silent Byzantine servers do not affect dramatically the
  performance.
\item (\Hdegrade) The performance does not degrade over time.
\end{compactitem}

%
%

To evaluate these hypotheses, we carried out the experiments described below and
reported in Fig.~\ref{fig:experiments}.
In all cases, operations are injected by clients running within the
same Docker container.
Resident memory was always enough such that in no experiment the
operating system needed to recur to disk swapping.
All the experiments consider deployments with 4, 7, or 10 server nodes,
and each running experiment reported is taken from the average of 10
executions.

We tested first how many epochs per minute our \setchain
  implementations can handle. In these runs, we did not add any
  element and we incremented the epoch rate to find out the smallest
  latency between an epoch and the subsequent one. We run it with 4, 7,
  and 10 nodes, with and without Byzantines servers.
  This is reported in Fig.~\ref{fig:experiments}(a).
  
In our second experiment, we estimated empirically how many
  elements per minute can be added using our four different
  implementations of \setchain (\AlgTwo and \AlgThree with and without
  aggregation), without any epoch increment.  This is reported in
  Fig.~\ref{fig:experiments}(b).  In this experiment \AlgTwo and \AlgThree
  perform similarly. With aggregation \AlgTwo and \AlgThree also perform
  similarly, but one order of magnitude better than without
  aggregation, confirming (\Hagr).  Putting together
  Fig.~\ref{fig:experiments}(a) and (b) one can conclude that sets are
  three orders of magnitude faster than epoch changes, confirming
  (\Hspeed).
  
  In our third experiment,
  we compare the performance of our implementations combining epoch increments
and insertion of elements.
We set the epoch rate at 1 epoch change per second and calculated the
maximum add ratio. The outcome is reported in
Fig.~\ref{fig:experiments}(c), which shows that \AlgThree outperforms
\AlgTwo. In fact, \AlgThreeSet even outperforms \AlgTwoSet by a factor
of roughly 5 for 4 nodes and by a factor of roughly 2 for 7 and 10 nodes.
\AlgThreeSet can handle 8x the elements added by \AlgThree for 4 nodes
and 30x for 7 and 10 nodes. The benefits of \AlgThreeSet over
\AlgThree increase as the number of nodes increase because
\AlgThreeSet avoids the broadcasting of elements which generates a
number of messages that is quadratic in the number of nodes in the
network. This experiment confirms (\Hbetteralg) and (\Hagr). The
difference between \AlgThree and \AlgTwo was not observable in the
previous experiment (without epoch changes) because the main
difference is in how servers proceed to collect elements to vote
during epoch changes.

The next experiment explores how silent Byzantine servers affect
  \AlgThreeSet.  We implement silent Byzantine servers and run for 4,7 and
  10 nodes with a an epoch change ratio of 1 per second, calculating
  the maximum add rate.
  This is reported in~Fig.~\ref{fig:experiments}(d).
  Silent byzantine servers degrade the speed for $4$ nodes as in this
  case the implementation considers the silent server very frequently
  in the validation phase, but it can be observed that this effect is
  much smaller for larger number of servers, validating (\Hbyzantine).
  
In the final experiment, we run 4 servers for a long time (30
  minutes) with an epoch ratio of 5 epochs per second and add
  requests to 50\% of the maximum rate.
  We compute the time elapsed between the moment in which the client
  requests an add and the moment at which the element is stamped.
  Fig.~\ref{fig:experiments}(e) and (f) show the maximum and average
  times for the elements inserted in the last second.
  In the case of \AlgThree, the worst case during the 30 minute experiment
  was around 8 seconds, but the majority of the elements were
  inserted within 1 sec or less.
  For \AlgThreeSet the maximum times were 5 seconds repeated in many
  occasions during the long run (5 seconds was the timeout to force a
  broadcast). This happens when an element fails to be inserted using
  the set consensus and ends up being broadcasted.
  In both cases the behavior does not degrade with long runs,
  confirming (\Hdegrade).

As final note, in all our experiments all elements are signed and they are
added to the \setchain along with their signature in order to simulate
a signed element (like a transaction request) in a Blockchain.
Therefore, each elements is composed of a constant amount of random
bytes, a public key to simulate a sending address and a signature.
The \setchain servers verify each element signature before accepting
it in the add request.
%
For the implementation, we leveraged the ed25519 crypto library of the
Golang standard library.

In summary, considering as baseline the throughput of epochs (which
internally performs a set consensus), implementing \setchain is
three orders of magnitude more performant than consensus.
Also, the performance of the  algorithms is resilient to silent Byzantine servers.

\section{Distributed Partial Order Objects (DPO)}
\label{sec:wrappers}

\begin{figure}[t!]
  \begin{algorithm}[H]
    \caption{\small Correct client protocol for DPO (for Alg.~\ref{DPO-alg-basb} and~\ref{DPO-alg-basb3}).}
    \label{DPO}
    \begin{algorithmic}[1]
      \Function{\<DPO>.\<Add>}{$e$}
      \State \<call> \<Add>\((e)\) in \(f+1\) different servers.
      \EndFunction
      \Function{\<DPO>.\<Get>}{~}
      \State \<call> \<Get>\(()\) at least \(3f+1\) different servers.
      \State \<wait> \(2f+1\) resp $s.(\THESET, \HISTORY,\<epoch>)$ 
      \State $S \leftarrow \{ e | e\in{}s.\THESET \text{ in at least } f+1 \text{ servers } s \}$
      \State $H \leftarrow \emptyset$
      \State $i \leftarrow 1$
      \State $\Servers \leftarrow \{ s : s.\<epoch> \geq i \}$
      \While {$\exists E: |\{s: s.\<history>(i) = E \}| \geq f+1$}~\label{DPO-bwhile}
        \State $H \leftarrow H \cup \{ \langle i, E \rangle \} $
        \State $\Servers \leftarrow \Servers \setminus \{ s : s.\<history>(i)\neq E\}$ 
        \State $\Servers \leftarrow \Servers\setminus \{s: s.\<epoch>=i\}$         
        \State $i \leftarrow i+1$
      \EndWhile~\label{DPO-ewhile}
      \State \<return> $(S, H, i-1)$
      \EndFunction
		    \Function{\<DPO>.\<EpochInc>}{$h$}
			\State \<call> \<EpochInc>\((h)\) in \(f+1\) different servers.
			\EndFunction
    \end{algorithmic}
  \end{algorithm}
  \vspace{-2em}
\end{figure}

The algorithms presented in Section~\ref{sec:implementation} and the
proofs in Section~\ref{sec:properties} consider the case of a correct
client contacting a correct server.
Obviuosly, client processes do not know if they are contacting a Byzantine or
correct process, so a client protocol is required to encapsulate the details of
the distributed system.
We describe now such a client protocol inspired by the one for
DSO~\cite{Cholvi2021BDSO}.
Implementing a correct client involves the exchange of several more
messages than contacting a single server with a request, so we later
describe a more efficient alternative.

The general idea of the client protocol is to interact with enough
servers to guarantee enough correct nodes are reached to ensure the
desired behavior.
The API has methods that wait for a result ($\<Get>$) and methods that
do not require a response ($\<EpochInc>$ and $\<Add>$).
Alg.~\ref{DPO} shows the client protocol, which will be executed by
correct clients.
To contact with at least one correct server, we need to send $f+1$
messages, as for $\<Add>(e)$ and $\<EpochInc>(h)$.
Note that each message may trigger different broadcasts.

The wrapper algorithm for function $\<Get>$ can be split in two parts.
First, the protocol contacts $3f+1$ nodes, and waits for at least
$2f+1$ responses, because $f$ Byzantine servers may refuse to respond.
The response from server $s$ is $(s.\<theset>,s.\<history>,s.\<epoch>)$.
%
The protocol then computes $S$ as those elements known to be
in $\<theset>$ by at least $f+1$ servers (which includes at least one
correct server).
%
To compute the history $H$,
the code goes incrementally epoch by epoch as long as
at least $f+1$ servers within the set $\Servers$ (which is initialized with
all the servers that responded with non-empty histories) agree on a set $E$ of
elements in epoch $i$.
Note that if at least $f+1$ servers agree that $E$ is the set of elements in
epoch $i$, then $E$ is indeed the set of stamped elements in epoch
$i$.
Then, in the loop we remove from $\Servers$ those servers that either do
not know more epochs or that reported something different than $E$.
Once this process ends, the sets $S$ and $H$, and the latest processed epoch
are returned.
Note that it is guaranteed that $\<history>\subseteq\<theset>$.

Finally, we present an alternative faster optimistic client assuming
that servers sign each epoch with a cryptographic signature.
%
%
Correct servers sign cryptographically a hash of the set
of elements in an epoch, and insert this hash in the \setchain as an
element.
Clients only perform \textbf{a single} \(\<Add>(e)\) request to one
server, hoping it will be a correct server.
After some time, the client invokes a $\<Get>$ from
\textbf{a single} server
(which again can be Byzantine) and check whether $e$ is
in some epoch
signed by (at least) $f+1$ servers, in which case the epoch is correct and $e$ has been
successfully inserted and stamped.
Note that this requires only one message per $\<Add>$ and one message per $\<Get>$.
Optimistic clients can repeat the process after a timeout if the contacted server did not respond.


\section{Concluding Remarks}\label{sec:discussion}

We presented a novel distributed data-type, called \setchain, that
implements a grow-only set with epochs. The data structure tolerates Byzantine server
nodes.
We provided a low-level specification of desirable properties of
\setchains and presented three distributed implementations, where the
most efficient one uses Byzantine Reliable Broadcast and RedBelly set
consensus.
We also showed an empirical evaluation that suggests that the
efficiency of the set (in terms of elements inserted) is three orders
of magnitud higher than consensus.

Future work includes developing the motivating applications listed in
the introduction, including (1) an implementation
of mempools as \setchain\!\!s for logging,
(2) an encrypted version of mempools to
prevent front-running, and
(3) an improved implementation of L2 optimistic rollups.
%
We will also study how blockchains with a synchronized \setchain as a
side-chain can make smart-contracts more efficient.
An important problem to solve is, of course, how to make clients of
the \setchain pay for the usage (even if a much smaller fee than the
blockchain itself).

There is also interesting future work from the point of view of
the foundations of distributed systems.
As Alg.~\ref{DPO-alg-basb3} shows, Byzantine tolerant building blocks
do not compose well from the point of view of efficiency. 
They have
to be overly pessimistic and cannot exploit the fact that, when blocks
are composed to build a correct node, both the client and the server of the
block are same and can hence be considered correct.
Studying this as a general principle is an interesting problem for
future work.
Also, it is easy to see that the Byzantine behavior at the level of
\setchain server nodes can be modeled by simple interactions with BRB
and SBC so one can build a (non-deterministic) implementation that
simulates the behavior of any Byzantine server node.
Such a modeling would greatly simplify formal proofs of Byzantine
tolerant data structures.


\bibliographystyle{abbrv}
\bibliography{bibfile}
\end{document}